\documentclass{epl}
\usepackage{epsfig}
\title{Putting hydrodynamic interactions to work: tagged particle
 separation} 
\author{J.\ L.\ Iguain\inst{1,2} \and  J.\ Kurchan\inst{3}}
\institute{
\inst{1} Laboratoire de Physique Th\'eorique  et Hautes Energies, Jussieu, 
5\`eme \'etage,  Tour 24, 4 Place Jussieu, 75005 Paris, France, 
Unité Mixte de Recherche CNRS UMR 7589 \\
\inst{2} Present address: D\'epartement de Physique et Groupe de Recherche 
en Physique et
Technologie des Couches Minces (GMC),
Universit\'e de Montr\'eal, Case Postale 6128, Succursale Centre-Ville,
Montr\'eal, Qu\'ebec H3H 3J7, Canada.\\
\inst{3}  P.M.M.H. Ecole Sup{\'e}rieure de Physique et Chimie Industrielles,
10, rue Vauquelin, 75231 Paris CEDEX 05,  France,
Unité Mixte de Recherche CNRS UMR 7636
}
\begin{document}
\today

\maketitle

%%%%%%%%%%%%%%%%%%%%%%%%%%%%%%%%%%%%%%%%%%%%%%%%%%%%%%%%%%%%%%%%%%%%%%%%%%%%%%%%%
\begin{abstract}
Separation of  magnetically tagged cells is 
performed by attaching 
 markers to a subset of cells in suspension and
applying fields to  pull from them in a variety of ways. 
The magnetic force is proportional to the field
 {\em gradient},  and 
the hydrodynamic interactions play  only a passive, adverse  role.
We propose  using
  a homogeneous rotating magnetic field only to make tagged
 particles rotate,
and then performing the actual separation by means of
 hydrodynamic interactions, which thus play an  active role. 
The method, which we explore here theoretically and
 by means of numerical simulations, 
lends itself naturally to sorting on large scales.
\end{abstract}

%%%%%%%%%%%%%%%%%%%%%%%%%%%%%%%%%%%%%%%%%%%%%%%%%%%%%%%%%%%%
%%%%%%%%%%%%%%%%%%%%%%%
The appearance of immunomagnetic beads ---
  superparamagnetic nano or
micro-particles attached to an antibody ---
 increased considerably the possibilities
for cell detection and isolation.
 Suitable antibodies can be chosen to bind the beads to 
the desired cells, which are then magnetically marked, leading to  
 a very rich tool for separation. 

Immunomagnetic separation has  become  very popular in the last
years. However, there has 
been more progress in the quality of beads themselves than in
the processing methods\cite{sci}.
(For details on existing techniques, see, for example, \cite{rojo,viola}).
In the  usual situation, 
normal and tagged cells are in suspension in a liquid
and, in  all the existing separation techniques,  
 magnetic field {\em gradients} are used  to drag the 
latter through the liquid. 
  The need to maximize gradients in order to improve efficiency
 imposes 
 reduced geometries and hence introduces 
practical  limitations which may turn out to be important in applications 
 with very large total number of cells. 
 This can in principle be remedied  using 
 stronger magnetic fields and bead moments,
 but then undesired 
 clustering due to magnetic attraction between tagged cells appears.
On the other hand, cell concentrations have to be kept low to prevent
normal cells from being dragged by tagged ones through  hydrodynamic
interactions.

The goal of this letter is to introduce a very different approach:
it consists of using the magnetic field {\em only}
 to make the tagged
 particles spin,
 and to use the resulting  hydrodynamic interactions
to effect the separation. We test this approach
 numerically,  the method of simulation
we used is described later in this text.
 Forced spinning can be produced
 by a spatially
{\em homogeneous}  rotating magnetic field and, as we shall see,
 interparticle forces
 do
not represent a drawback but actually produce  the separation.
We shall consider two settings: The first, used for 
 continuous sorting
of particles, is reminiscent of  Free-Flow
 Magetophoresis\cite{rojo}, although it may be implemented in a wide channel. 
The second is a many-body dynamical phase-separation transition,
 in itself
and intriguing phenomenon that to our knowledge has not been described before.

\begin{figure}[h]
\centerline{
\psfig{file=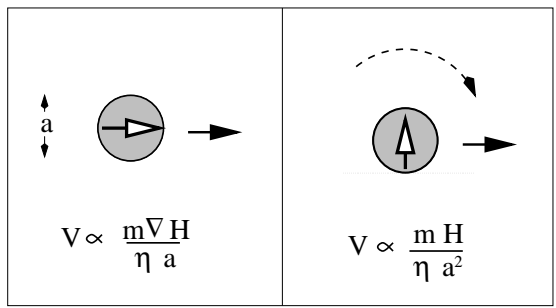,width=4cm,angle=-90}
}
\vspace{.5cm}
\caption{A particle being pulled by a force gradient (left), and made
  to roll on a rough line by a rotating field (right). The black and
  white arrows
  indicate the direction of the field and moment, respectively (both
  rotate in the figure on the right).}
\label{Fig0}
\end{figure}

Let us first discuss why it is avantageous to use
 fields to {\em spin}, rather than to {\em displace}
 the 
particles.
Consider a spherical particle of radius $a$ and magnetic dipole moment
 $m$  inmersed in a fluid of viscosity
$\eta$. We can displace   the particle by applying a static field
 Fig.\ \ref{Fig0} (left), in which case the force is proportional to the field's
{\em gradient}.  Alternatively, we 
 can apply  a rotating field (right), and the torque and  angular
 velocity will be proportional to the field {\em amplitude}. In order to
 convert rotation into translation we need an extra force: in this
 simple example we suppose we can force the particle to rotate without
 sliding on a line.

Let us compare the efficiency of each method:
 Applying  a  magnetic field $H$   the  force on the particle is
 $ \propto m\; \nabla H \sim m\;{H}/{\ell}$, where $\ell$
is the typical length over which the field varies. The resulting
 velocity
 is given by Stokes' law
$v_{static} \sim \frac{m\; H}{6 \pi \eta a \ell}$.
 If instead we apply a  rotating field, setting the field's frequency $\omega$  
to be the fastest that  the particle's rotation can follow,
the angle between field and magnetic moment
approaches ninety degrees, and the torque $\tau$ on the particle
is $\sim m H$. For an isolated particle, Stokes' law gives
$\omega \sim \frac{m\;  H}{8 \pi \eta a^3}$. 
Imposing that the particle
 rolls without sliding on the line within  the liquid, 
the rotational motion  is converted by friction into  linear
motion $v_{rot}= wa$, i.e.
$
v_{rot} \sim a \frac{m\;  H}{8 \pi \eta a^3} \propto \frac{m\;  H}{\eta
  a^2}$.
Comparing both mechanisms, we have:
  \begin{equation}
\frac{v_{rot}}{v_{static}} \propto \frac{H}{\nabla H \;  a } \sim 
\frac{\ell}{a} 
\label{jj}
\end{equation}
{\em i.e. the advantage of using rotation over
  using  a static field is in  proportion
to  the typical range of variation of
fields (e.g. the distance between magnetic poles), 
compared with the particle size --- typically a few orders of 
magnitude~\cite{uno}}.

  In what follows, we will show that one
  can apply the same principle~\cite{MIXSEP},
 but with the role of transforming
  rotational  into translational motion played by the  
 hydrodynamic interactions, and to a lesser extent, the direct
inter-particle forces. Depending on the setting, the interaction is 
 {\em i)} between the rotating cells and a porous medium or a specially
designed set of obstacles, {\em ii)}  between marked and  unmarked
 cells and 
 {\em iii)} between cells  and the walls.

Let us note here that the ferromagnetic particle need not
  have a permanent  magnetic
moment: the only necessary condition is that when subjected to a rotating
 field the angle of the magnetization direction
 lags with respect to that  of the field
(i.e. that there is hysteresis), and this will happen in a ferromagnetic
substance whether permanently magnetized or not. In the latter case
the 
estimate
(\ref{jj}) will be modified  by a hysteresis-dependent factor,
but the force will still depend upon  the field rather than its
gradient. 

Hydrodynamic mechanisms that convert rotation into lateral displacements
 have been known for a long time (for example, see \cite{Se,si}): 
the question here is to devise a method that works at zero Reynolds
number so it does not
become inefficient  for 
particle sizes and liquid viscosities
 involved in biological applications. 

\vspace{.2cm}

{\bf Stokesian Dynamics}

\vspace{.2cm}

Let us briefly discuss how the results  in this letter where obtained.
The dynamics of the particles in the suspension at zero Reynolds
number
was simulated
by means of a Stokesian Dynamics (SD) algorithm
\cite{Brady,Brady1,obrien,beena,jeffrey,kim,Brady2,Brady3},
which  involves various levels of approximation.
We use the so-called F-T (force-torque) version which provides
excellent results when no external linear shear is imposed on
the flow and diminishes the  computing time by a factor
$\approx 6.2$ with respect
to the most elaborate approximation\cite{Brady}.
At this level, the velocities of the $N$ suspended particles and the
forces and torques
acting on them are related by a grand resistance matrix
$\cal R^*$, $F$= $\cal
 R^*$($U-\langle u \rangle$),
where the $6N$-dimensional  vectors $F$ and $U$ represent the
applied forces-torques and the
linear-angular velocities, respectively, and $\langle U \rangle$
is the average velocity of the
suspension (particles and fluid).
Details of the grand resistance matrix construction
are given in ref \cite{Brady1}.
Briefly, it works in two steps: The inverse of $\cal R*$ is first
approximated by a far-field, multipole expansion. The box of
containing the
$N$ particles is periodically repeated \cite{obrien} and the
convergence of the sum over all the long-range
interactions
is accelerated by the Ewald summation technique\cite{beena}.
Near field contributions are finally included in a pairwise
additive fashion at the level of $\cal R^*$ using exact two-sphere resistance
functions\cite{jeffrey,kim}.

In all our SD simulations, every particle has a radius $a$.
The elementary cell
containing the $N$ suspended particles is
 a box of sides  $L_x$, $L_y$ and $L_z$
in the directions $x$, $y$ and $z$, respectively.
The particles are always in the
 $x-y$ plane.
Applied torques are in the $z$ direction.
The obstacles used for continous separation are represented by immobile particles
which have the same radius that the suspended ones,
and similarly,
the bottom wall we need for collective segregation is simulated
with fixed particles of radius $a$ (see for example in ref.\cite{Brady2,Brady3}).
Finally, sedimentation, when present (see Fig.\ \ref{Fig1}), is in the
negative $y$ direction.

\vspace{.2cm}

{\bf  Continuous separation.}

\vspace{.2cm}

An implementation of 
continuous sorting or 
enriching of tagged cells that works in the high viscosity limit is
the following: 
 An average  flow  of velocity  $U_d$ is sustained
through a medium with fixed obstacles.   These could for example be
 filaments 
 with axis perpendicular 
to the flow  --- like the hairs of
a brush. A rotating field is applied, imposing that marked particles rotate 
(with angular velocity $\omega$, when isolated)
around an axis parallel to the obstacles.

\begin{figure}[h]
\centerline{
\hspace*{.5cm}
\psfig{file=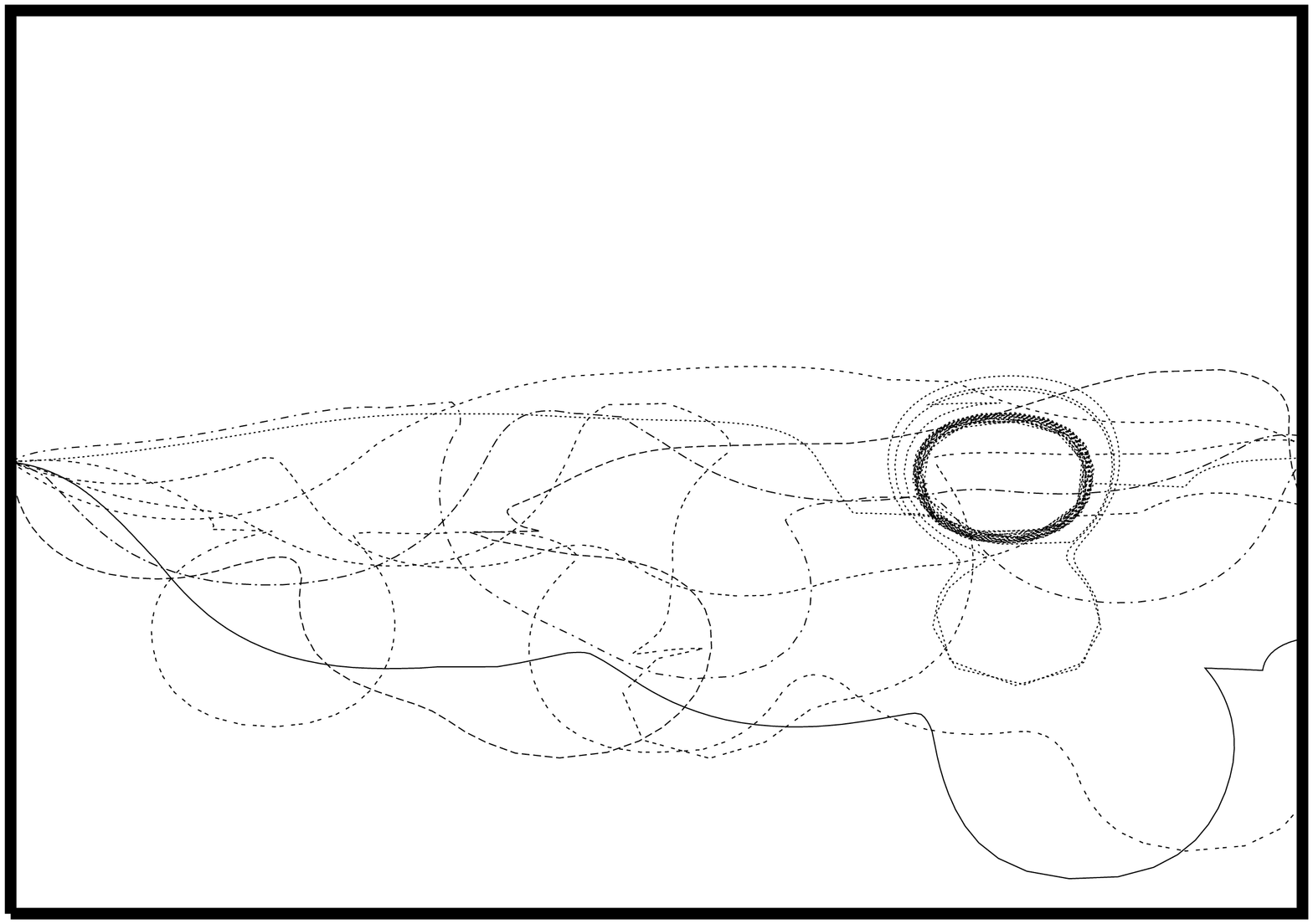,width=5cm,height=5cm,angle=-90}
\psfig{file=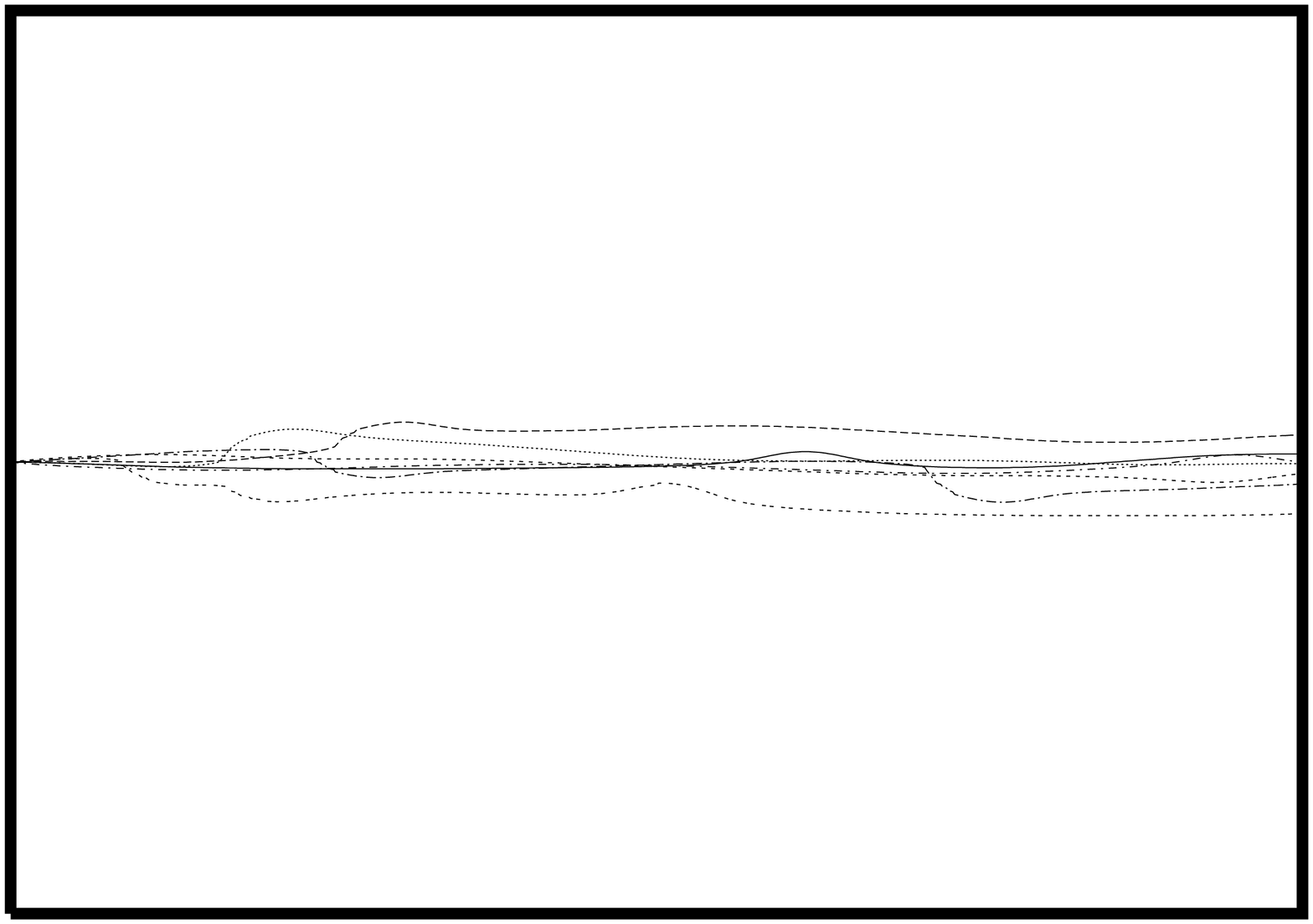,width=5cm,height=5cm,angle=-90}
\vspace{.5cm}\\
}
\caption{ Stokesian dynamic trajectories corresponding to a particle of
radius $a$ entering
 the fluid on the left. Fourteen fixed obstacles, also  of radius $a$
were placed
at random  in the  rectangle (not shown).
 Each curve corresponds to a different
configuration  of the  obstacles. Left: marked particles 
under a torque $\tau$ perpendicular to the paper and $\omega a/  U_d =
100$. Right: unmarked particles. 
 The sides of the unit cell are
$L_x=40 a$, $L_y=40 a$ and $L_z=200 a$. }
\label{coins}
\end{figure}

\begin{figure}[h]
\centerline{
\psfig{file=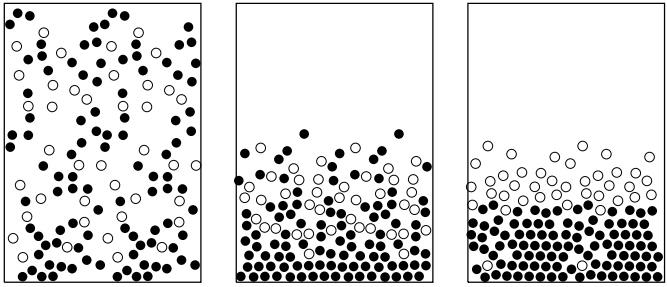,width=9cm,angle=-0}
}
\vspace{.5cm}
\caption{Segregation of marked (white) and unmarked (black)
 particles in
a Stokesian dynamics simulation under sedimentation of every
particle and forced spinning of the white particles.
A bottom wall is simulated by a string
of fixed particles (not shown).
$n' = 1000$.
 The time-scales (in $2\pi/\omega$)
 are, 300 (left), 3000 (center) and 30000 (right). The unit cell
has dimensions $L_x=20 a$, $L_y=200 a$ and $L_z=200 a$ (two shown).}
\label{Fig1}
\end{figure}

 Particles forced to rotate are made by hydrodynamic
interactions
to describe arcs around the obstacles: if their density  is high enough  that
these deflections overlap,  a lateral diffusivity results
(Fig.\ \ref{coins}-(left)).
 On the contrary, unmarked particles will be carried by the flow, gently
avoiding the obstacles  (Fig.\ \ref{coins}-(right)).
A continuous enrichment method can  then be implemented by
selecting through appropriate windows (at the right hand side of the channel shown in  Fig.\
\ref{coins})
either the magnetic or the non-magnetic 
particles.

 Purely
hydrodynamic interactions give a symmetric diffusivity on
 average~\cite{dos}. Although unnecessary for enrichment purposes,
 one can also generate a net lateral
 drift. Marked-particle diffusivity increases 
with obstacle density:
then, by making the latter spatially dependent, one can take advantage of the
general fact that diffusive particles
 tend to accumulate in  regions of smaller diffusivity \cite{Risken}.

In the usual free-flow magnetophoresis \cite{FFM} a pair of
wedge-type magnetic poles on the sides of a  channel that necessarily 
has to be thin, in order to maximize the field gradient.
In the method described above,  the separation is proportional
to the intensity of the rotating field, there is no need to have a field gradient 
and hence there is the possibility of working with a thicker
channel, weaker fields or smaller marker moments.
Let us make a rough comparison:
When the deflection is caused by a stationary field, we may 
give a particle a lateral velocity $U_{trans}$. The separation will be 
the faster, the larger the drift velocity $U_d$, as this reduces the passage
 time $t_{trans} \sim L/U_d$, with $L$  the typical length of the 
channel. However, geometry requires that particles deflect a certain angle 
$\tan\theta=U_{trans}/U_d$, so that $t_{trans} \sim L \tan 
\theta/U_{trans}$.
On the other hand, the deflection in a system like Fig.\ \ref{coins}
is a complicated function $f(n)$ of the adimensional number $n=\omega a/U_d$,
where $\omega$ is the angular velocity of a single tagged particle under the
rotating  field, and we have the requirement that $\tan \theta = f(n)$.

 Using Stokes' law for rotation and translation, in an entirely
analogous way to that leading to Eq. (1), 
if we compare devices working with either principle
having the same  $\theta$, $L$ and $H$, 
and working with tagged particles with the same  $m,a$,
we have:
 \begin{equation}
\frac{t_{trans}}{t_{rot}}  \propto
 \frac{H}{a \nabla H} \sim \frac{\ell}{a}
\label{siete}
\end{equation} 
which is precisely the estimate in the simple example of the rolling particle.

\vspace{.2cm}

{\bf Collective segregation}
\vspace{.2cm}

The second method involves no obstacles, and uses
phase-separation under gravity. It may in principle be used to enrich
arbitrarily low concentrations of marked particles.

 Collective segregation results when all the particles (marked and unmarked)
sediment under a rotating magnetic field. In such circumstances,
 isolated particles fall with velocity  $U_g=F_s/(6\pi\eta a)$
(where $F_s$ is the sedimentation force)
and spin with angular speed $\omega$
if they are marked. $U_g$ can be increased by centrifugation, and decreased by matching the
density of the
particles  to that of the fluid. 

The basic adimensional 
parameter of the problem is the ratio of speed of rotation of an
isolated  tagged particle $\omega$  to the speed of 
sedimentation  $U_g$ of an isolated particle 
$n'= \frac{\omega a}{U_g}$ which
 can be made large either by increasing the field (and hence
$\omega$), or by reducing the sedimentation velocity $U_g$.
{\em At  given geometry and $n'$},
the time taken for separation depends on
$\omega$, and hence can be measured in particle revolutions.

In Fig.\ \ref{Fig1} we show how separation of tagged (white) 
and normal (black) particles happens. Initially
all the particles are randomly distributed. A sedimentation force
 $F_s$ acts on every particle
in the negative vertical direction while a torque $\tau$, perpendicular
 to the paper, is applied 
only to the white particles. The corresponding value of $n'$  is 
$n'=1000$ and the times, from
left to right are 300, 3000 and 30000  revolutions 
$2\pi/\omega$. Separation
is  achieved in a few  thousand particle revolutions.
Note that there is in principle no limitation on the width of the
particle bed, and that the method works with arbitrarily small
quantities of marked particles

The reason for this separation is twofold.
 Cells interact with their neighbours through 
hydrodynamic interactions. Those that are forced to rotate
   perturb  their near-neighbours, forming a local
 region in which particles diffuse strongly 
 --- {\em even if the thermal Brownian motion itself
 is negligible}. Because this diffusion depends on the presence of neighbours,
it is weaker in regions of smaller density:
this creates a net motion of the rotating particles 
from regions of high to regions of low
 particle density --- in this example, the surface. 
This is like having a can of living and dead ants: if the living ones diffuse
randomly  stepping on the dead bodies, they will eventually reach the surface.

\begin{figure}[h]
\centerline{
\psfig{file=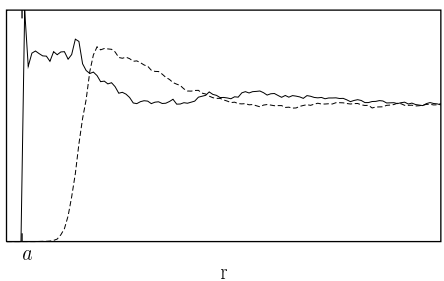,width=7cm,angle=-0}
}
\vspace{.5cm}
\caption{ Expulsion of neighbours effect. The plots show the
distribution function of distances between 
unmarked particles (full line), and between marked and other particles
 (dashed line). The curves were calculated with a long-time Stokesian dynamics 
simulation
with 15 free  particles and 15 particles under a torque $\tau$, 
in a cubic cell of side $100 a$.}
\label{Fig2}
\end{figure}

 A second, competing effect is the following: the rotating particles also
  tend
to expel their neighbours
 (see Fig.\ \ref{Fig2}). Hence, they create a low-density
region that tends to migrate to the surface through buoyancy. 

This separation method is strongly reminiscent of the `Brazil nut'
phenomenon
---
  the fact that larger particles of a vibrated granular medium
tend to go to the surface ---
 since in  some cases this can be attributed to the
fact that  larger particles tend
to have stronger diffusivity \cite{leo}: here this increased diffusivity is
 explicitly induced in the
marked particles by the rotating field.

\vspace{.2cm}

{\bf Robustness}

\vspace{.2cm}

We have 
described a separation technique whose basic mechanism  is the diffusion
generated by the particles that are forced to rotate continuously or
alternatively, through the interaction with their   
 closest neighbours or with nearby obstacles.
This diffusion is  used to sort the particles, by non-magnetic means. 
The relevant physical principle    is robust: the detailed 
motion the magnetic field induces on the particle,
 as well as the precise form of the
interactions
 are not of crucial importance.
Indeed, we have performed a molecular dynamic simulation with 
particles interacting through different kinds of  forces 
and confirmed that rotating particles still separate.
Furthermore,  although we have assumed that the tagged particles
rotate around their centers, the basic behaviour would be the same even if only
the magnetic 
tags rotate, since that would in itself create the necessary hydrodynamic 
interactions.

 The method  allows  to use smaller fields, thus avoiding 
 the magnetic interactions between particles, that may
 cause an unwanted clustering.
 It may also be used with  larger volumes, and to
  enrich samples with a very small concentrations of tagged
 particles.  Although we have concentrated mostly on cell separation, the
basic mechanism should apply to other kinds of particles.

We wish to thank  M. Fermigier, F. Feuillebois, M. Martin, J-E. Wesfreid and
specially M. Hoyos for useful discussions.

%%%%%%%%%%%%%%%%%%%%%%%%%%%%%%%%%%%%%%%%%%%%%%%%%%%%%%%%%%%%%%%%%%%%%%%
%                    BIBLIOGRAPHY
%%%%%%%%%%%%%%%%%%%%%%%%%%%%%%%%%%%%%%%%%%%%%%%%%%%%%%%%%%%%%%%%%%%%%%%


\begin{thebibliography}{99}
\bibitem{uno}Depending on the setup, $\ell$ is of the order of  a  tenth of a
millimeter (in random arrays of steel wire) to a few millimeters in a
  flowing chamber.
\bibitem{dos}The slight net downward drift in Fig.\ \ref{coins},
 is a consequence of the
the core repulsive (non-hydrodynamic) force between particles.
\end{thebibliography}
\end{document}